# Crystal structure and high-field magnetism of $La_2CuO_4$

M. Reehuis [1,2], C. Ulrich [1], K. Prokes [2], A. Gozar[3], G. Blumberg [3], Seiki Komiya[4], Yoichi Ando[4], P. Pattison[5,6], and B. Keimer [1]

[1] *Max-Planck-Institut für Festkörperforschung, Heisenbergstr. 1, D-70569 Stuttgart, Germany*
[2] *Hahn-Meitner-Institut, Glienicker Str. 100, D-14109 Berlin, Germany*
[3] *Bell Laboratories Lucent Technologies, Murray Hill, New Jersey 07974, USA*
[4] *Central Research Institute of Electric Power Industry, 2-11-1 Iwato-kita, Komae, Tokyo 201-8511 Japan*
[5] *SNBL at ESRF, BP 220, F-38042 Grenoble Cedex 9, France*
[6] *Laboratory of Crystallography, Swiss Federal Institute of Technology, BSP-Dorigny, CH-1015 Lausanne, Switzerland*

Neutron diffraction was used to determine the crystal structure and magnetic ordering pattern of a $La_2CuO_4$ single crystal, with and without applied magnetic field. A previously unreported, subtle monoclinic distortion of the crystal structure away from the orthorhombic space group *Bmab* was detected. The distortion is also present in lightly Sr-doped crystals. A refinement of the crystal structure shows that the deviation from orthorhombic symmetry is predominantly determined by displacements of the apical oxygen atoms. An in-plane magnetic field is observed to drive a continuous reorientation of the copper spins from the orthorhombic *b*-axis to the *c*-axis, directly confirming predictions based on prior magnetoresistance and Raman scattering experiments. A spin-flop transition induced by a *c*-axis oriented field previously reported for non-stoichiometric $La_2CuO_4$ is also observed, but the transition field (11.5 T) is significantly larger than that in the previous work.
PACS numbers: 61.12.Ld, 61.50.Ks, 75.25.+z,

# I. INTRODUCTION

La$_{2-x}$Sr$_x$CuO$_4$ is one of the simplest high temperature superconductors, and there has been a large amount of work on this system since the discovery of high temperature superconductivity. Some of the most incisive information has come from neutron scattering, which has provided a detailed microscopic picture of the evolution of the structural and magnetic properties as a function of doping. [1] The parent compound of this family, undoped La$_2$CuO$_4$, is a Mott insulator that orders antiferromagnetically with a Néel temperature, $T_N$, of approximately 325 K. [2] Antiferromagnetic long-range order vanishes for $x \approx 0.02$ and is replaced by phases with short-range magnetic order that partially coexist with superconductivity. Neutron scattering data on La$_{2-x}$Sr$_x$CuO$_4$ have provided the major impetus and motivation for one of the most prominent microscopic models of high temperature superconductivity, according to which superconductivity is seeded in the charge 'stripes' that can be either static (in underdoped materials, in particular around the commensurate hole content $x = 1/8$), or fluctuating (in optimally doped and overdoped materials). [1,3]

La$_{2-x}$Sr$_x$CuO$_4$ exhibits a tetragonal crystal structure (with space group *I4/mmm*) at high temperatures, which transforms into an orthorhombic structure (with space group *Bmab*) below a transition temperature up to $T_S = 520$ K, that strongly depends on doping. [4-6] In the orthorhombic phase, macroscopic samples invariably exhibit crystallographic twinning (that is, the formation of micron-scale domains in which the *a*- and *b*-axes in the copper oxide planes are interchanged), unless the sample is cooled through $T_S$ under uniaxial stress. Most experimental work on La$_{2-x}$Sr$_x$CuO$_4$, including almost all neutron scattering work, has thus been performed on twinned samples.

Recent developments call for a reinvestigation of the interplay between the magnetic and structural correlations in untwinned La$_{2-x}$Sr$_x$CuO$_4$. First, the charge dynamics [7] and the spin susceptibility [8] of lightly doped La$_{2-x}$Sr$_x$CuO$_4$ were found to exhibit sizable in-plane anisotropies, possibly as a consequence of stripe formation. A quantitative correlation with the in-plane anisotropy of the spin correlations is desirable. Second, it was recently shown that it is possible to detwin pure and lightly doped La$_2$CuO$_4$ crystals by applying a magnetic field at room temperature. [9] This indicates a novel magneto-structural coupling mechanism that is thus far not understood. Third, magnetoresistance measurements [10] and Raman scattering [11] experiments in high magnetic fields have indicated a reorientation of the magnetic moments and associated transport anomalies when a magnetic field is applied along the orthorhombic *b*-axis in the copper-oxide layers. These results, as well as corresponding theoretical work [12], have yielded predictions for the magnetic field dependence of magnetic Bragg reflections observable in neutron scattering experiments. These predictions partially contradict earlier work. [13]

Motivated by these results, we have studied the crystal and magnetic structure of an untwinned $La_2CuO_4$ single crystal by neutron diffraction. We find a previously unknown, subtle monoclinic distortion of the crystal structure resulting predominantly from displacements of the apical oxygen atoms. This distortion is also present in lightly Sr-doped $La_2CuO_4$ and may thus be relevant for a complete understanding of stripe formation and charge transport in this system. We also find two magnetic field-induced spin reorientation transitions. When the field is applied along the *c*-axis, we reproduce a previously observed [14] spin-flop transition, albeit at a significantly higher field than that reported earlier. An in-plane magnetic field induces a gradual reorientation of the magnetic moments from the *b*- to the *c*-axis. This confirms the predictions made on the basis of the Raman scattering and magnetoresistance work [10-12].

## II. EXPERIMENTAL DETAILS

High quality single crystals of $La_{2-x}Sr_xCuO_4$ were grown by the travelling-solvent floating-zone technique as described earlier. [8] For the experiments we used untwinned crystals of $La_2CuO_4$, $La_{1.99}Sr_{0.01}CuO_4$ and $La_{1.97}Sr_{0.03}CuO_4$ with dimensions $1.45 \times 1.45 \times 2.60$ mm$^3$, $2.10 \times 1.80 \times 2.75$ mm$^3$ and $1.60 \times 2.10 \times 2.55$ mm$^3$, respectively.

Neutron diffraction experiments were carried out on the four-circle diffractometer E5 at the BER II reactor of the Hahn-Meitner-Institut in Berlin. This instrument uses Cu and pyrolytic graphite (PG) monochromators selecting the neutron wavelengths 0.884 Å and 2.39 Å, respectively. The samples were placed in a strain-free mount inside a closed-cycle He cryostat. The data were collected with a two-dimensional position sensitive $^3$He-detector, $90 \times 90$ mm ($32 \times 32$ pixels). The refinements of the crystal and magnetic structures were carried out with the programs *Xtal* 3.4 and *FullProf*, respectively. [15,16] Here the nuclear scattering lengths $b(O) = 5.805$ fm, $b(Cu) = 7.718$ fm, $b(Sr) = 7.02$ fm, and $b(La) = 8.24$ fm were used. [17] For the determination of the magnetic order of the copper sublattice, the magnetic form factor of the $Cu^{2+}$-ion was taken from Ref. 18.

Elastic neutron scattering experiments in magnetic fields up to 14.5 T were performed in a vertical-field cryomagnet installed at the two-axis diffractometer E4 at the Hahn-Meitner-Institut. A PG monochromator working with the (200) reflection was used to select an incident wavelength of 2.44 Å. In order to reduce the contributions from second-order Bragg reflections, a PG filter was placed into the neutron beam.

In a further effort to clarify the evidence of a fine splitting of reflections resulting in a lower lattice symmetry of $La_2CuO_4$ high-resolution powder diffraction studies using synchrotron radiation were performed. The experiments were carried out with the high-resolution powder diffractometer on the BM1B (Swiss-Norwegian) beam line at the European Synchrotron Research Facility (ESRF) with photon wavelength 0.5200 Å. The sample was filled into a 0.5 mm diameter capillary and it was cooled with a cryostream nitrogen cooling system. A complete powder pattern with a $2\theta$ range between 1 ° and 55 ° was collected at 100 K. The Rietveld refinements of the powder diffraction data were carried out with the program *FullProf*.[16]

The degree of twinning in the crystals was determined by neutron diffraction from the ratios of the intensities of the reflections *h*0*l* and 0*hl*. As we will discuss in Section III.A, the reflections 0*hl* in the orthorhombic (*Bmab*) and in the monoclinic (*B*2/*m*) structure are limited to those for which *l* is an even number, whereas the reflections *h*0*l* are systematically absent with any odd number. The ratio $I(h0l)/I(0kl)$, where *h* is an odd and *l* an even number, therefore describes the degree of twinning. The relatively strong reflections 032 and 052 are well suited for this purpose. In the case of $La_2CuO_4$ the small ratio of 0.006 showed that the crystal is almost completely detwinned. In contrast, for $La_{1.99}Sr_{0.01}CuO_4$ and $La_{1.97}Sr_{0.03}CuO_4$ ratios of about 0.10 were found, indicating that the crystals are partially twinned.

### III. RESULTS AND DISCUSSION

#### A. Crystal structure

In order to refine the crystal structure of $La_2CuO_4$, a data set with a total number of 1024 (481 unique) reflections was collected at 325 K (that is, above the Néel temperature of this sample $T_N$ = 316(2) K). Refinements were carried out initially in the orthorhombic space group *Bmab* (*B* 2/*m* $2_1$/*a* 2/*b*). For the extinction correction the formalism of Zachariasen (type I) was used. The refinable parameter *g* in this formalism is related to the mosaic distribution, and the refinement resulted in the value of *g* = 608(3) rad$^{-1}$. For the absorption correction, the absorption coefficient $\mu$ = 0.0734 mm$^{-1}$ was used. The refinement of the overall scale factor as well as the positional and anisotropic thermal parameters resulted in a residual $R_F = 0.019$ ($R_w = 0.021$). Additionally the occupancies of the different atoms were refined, where the overall scale factor, the $g$-value, the positional parameters and the temperature factors were fixed. The results of the refinements are summarized in Tables I and II. Fig. 1 shows a pictorial representation of the crystal structure of $La_2CuO_4$. Here it can be seen that the thermal atomic vibrations of both the O1- and O2-atoms are enhanced as a consequence of soft octahedral tilting modes known from prior work. [19] The thermal parameters do not change significantly in the refinements using the lower-symmetric monoclinic space groups described below.

In order to check for a possible lower symmetry of the crystal structure, we measured reflections that are forbidden in *Bmab*. For all samples, we found additional, very low intensities (about 3 orders of magnitude smaller than the strongest reflections) on $0k0$ with $k = $ odd, $hk0$ with $h = $ even, $k = $ odd and $h0l$ with $h, l = $ odd. These conditions imply the absence of the glide planes *a* and *b*. This means that the direction of translation of the atoms along the *a*- and *b*-axes is no longer necessarily $t = a/2$ and $t = b/2$. Due to the lower symmetry one can expect an additional distortion on the $CuO_6$-octahedra. On the other hand the reflections are limited by the condition $h + l = 2n$. This implies the presence of a *B*-centred lattice. The highest-symmetry space group compatible with these constraints is the monoclinic space group *B*2/*m* (*B* 2/*m* 1 1). In this structure, the octahedral distortions in all $CuO_2$ planes are identical due to the translation $t = (a + c)/2$.

Our structure refinements in the space group *B*2/*m* showed that the intensities of the 45 reflections $0k0$ ($k = $ odd) and $hk0$ with $h = $ even were calculated much smaller than the corresponding observations. Therefore we carried out refinements in the lower-symmetric noncentrosymmetric space groups *B*2 (*B* 2 1 1) and *Bm* (*B m* 1 1). The better fit resulted from a refinement in *Bm* with $R_F = 0.027$ in comparison to $R_F = 0.030$ from the refinement in *B*2. Here it has to be mentioned that some of the refinable parameters were highly correlated, with correlation coefficients exceeding 90 %. (The correlation coefficient between two parameters $p_i$ and $p_j$ is obtained from the covariance, i.e. the second moment with respect to $p_i$ and $p_j$ by normalization $corr(p_i, p_j) = cov(p_i, p_j)/[\sigma(p_i) \sigma(p_i)]$, where $\sigma$ denotes the standard deviation). Therefore we used constraints restricting the values of particular positional parameters. This has been done for the positional parameters of the lanthanum atoms, since they should not be influenced by additional distortions of the $CuO_6$-octahedra. Further constraints of the $z$-parameters were applied for the other atoms. Attempts of systematic refinements showed that the loss of the translational component of ½ along *b* for the apical oxygen atoms (in *Bm*: 4 atoms, each in the Wyckoff position 2*a*) has the strongest influence on the intensities of the reflections forbidden in *Bmab*. On the other hand, shifts of the copper atoms (2 atoms, each in 2*a*) and the in-plane oxygen atoms (2 atoms, each in 4*b*) seem to be weak. The results of the refinement are summarized in Table I. In Table III it can be seen that the agreement between the observed and calculated structure factors is reasonable, albeit not fully satisfactory. A more precise determination of the atomic positions is prevented by high correlations between the parameters, as discussed above.

In order to investigate the monoclinic distortion of the Bravais lattice, we have further collected high-resolution synchrotron powder diffraction data of $La_2CuO_4$ at 100 K. The presence of a monoclinic lattice should be clearly manifested by a splitting of particular reflections. Rietveld refinements were carried out in the monoclinic space group *B*2/*m*. Our refinements finally did not show any evidence of a monoclinic distortion, but from our data analysis we were able to place upper bounds of 0.02° on possible deviations of $\alpha$, $\beta$, and $\gamma$ from 90°.

Structure refinements were also carried out on the doped compounds $La_{1.99}Sr_{0.01}CuO_4$ and $La_{1.97}Sr_{0.03}CuO_4$ from data sets collected at room temperature with a total of 1359 (472 unique) and 1056 (460 unique) reflections, respectively. The refined atomic parameters within *Bmab* were in a good agreement with those obtained for the pure compound given in Table I. Additional reflections indicating monoclinic distortions of magnitude comparable to that of $La_2CuO_4$ were also detected in both compounds. Therefore we can conclude that the deviation from the previously reported orthorhombic symmetry, i.e. a monoclinic distortion with the *Bm* space group, is intrinsic for pure and lightly doped $La_{2-x}Sr_xCuO_4$.

### B. Magnetic structure

In view of the complex crystal structure discussed above and the magnetic field induced spin reorientations to be discussed below, we begin this section with a brief review of possible magnetic structures in the framework of Bertaut's representation analysis.[21,22] Within the space group *Bmab* the copper atoms are located at the Wyckoff position 4*a* in the point symmetry 2/*m*•• with the four positions 0,0,0 (1), 0,½,½ (2), ½,0,½ (3) and ½,½,0 (4). The basis functions of the spin sequences at these positions are denoted as follows: *F*(+ + + +), *G*(+ − + −), *C*(+ + − −), and *A*(+ − − +). Table V lists the compatibility relations between these modes resulting from the representation analysis.

For completeness, we also show the results of the representation analysis in the monoclinic space group $B2/m$ in Table VI. Here one finds two different copper atoms at the Wyckoff position $2a$ (in 0,0,0; ½,0,½ ) and $2d$ [in ½,½,0; 0,½,½]. The two magnetic moments of each site can couple either ferromagnetically or antiferromagnetically as designated with the notations $A(+ -)$ and $F(+ +)$. The two copper sublattices are decoupled, and their type of magnetic order must not necessarily be the same. However, the results presented in section III. A indicate that the monoclinic distortion is very weak, so that its effect on the magnetic structure is almost certainly negligible. Therefore we hence present our data in the framework of the space group $Bmab$.

For $La_2CuO_4$ the strongest magnetic intensity was observed at the position of the reflection 100, in agreement with prior work. [2,22] In Fig. 2 it can be seen that the intensity of the reflection 100 vanishes at the Néel temperature $T_N$ = 316(2) K. This is in good agreement with the values $T_N$ = 325 K and $T_N$ = 310 K of fully oxygenated samples reported earlier. [2,11] This suggests that the magnetic moments of the copper atoms are ferromagnetically aligned in the $bc$-plane. Furthermore, as discussed below, the coupling between the copper moments is purely antiferromagnetic along the $x$-direction. Our refinements showed that the magnetic moments are aligned parallel to the $b$-axis. This is also in agreement with the magnetic structure presented earlier. [2,22,23] In the notation introduced above, the primary magnetic mode is thus $C_y$.

Reflections corresponding to $G$- or $A$-modes were not detected. If $C_x$- or $C_z$-modes (which are allowed in $B2/m$) were present, magnetic intensity should appear on the position of the reflection 010. We have seen above that weak intensity of structural origin is present at this position. At lower temperature, the intensity at 010 increases slightly. If the difference between the structure factors at 10 K and 325 K is attributed to magnetic ordering, it would correspond to a small moment of 0.060 $\mu_B$ along the $a$-axis. However, the more likely origin of the difference in 010 is not a magnetic but a nuclear effect, i.e. a slight increase of the monoclinic distortion with decreasing temperature. This confirms that the magnetic coupling along the $x$-direction is antiferromagnetic, which results in a $C_y$ magnetic structure. It can be seen in Table V that the mode $C_y$ is compatible with the mode $A_z$. This mode was not detected directly in our neutron diffraction experiments, but a weak $A_z$ contribution has been inferred from an analysis of the uniform susceptibility. [24] The full magnetic structure including both modes is shown in Fig. 3. From the refinement of five strong magnetic reflections at 10 K, we obtained a sublattice magnetization of 0.42(1) $\mu_B$, resulting in a residual $R_F$(mag) = 0.027. We note that this value was derived based on the isotropic Cu form factor of Ref. 18. However, the value will not be modified substantially if the anisotropic form factor of the Cu $d_{x^2-y^2}$-orbital is used. [25]

Our $La_{1.99}Sr_{0.01}CuO_4$ crystal exhibits a Néel temperature of 227(2) K (Fig. 2). The magnetic structure is identical to that of $La_2CuO_4$, where the saturated sublattice magnetization is 0.35(2) $\mu_B$. For $La_{1.97}Sr_{0.03}CuO_4$, no magnetic reflections were detected, consistent with the magnetic phase diagram reported earlier. [1,6]

### C. Magnetic field dependent measurements

We now turn to the magnetic field-dependent experiments, beginning with fields applied parallel to the copper-oxide planes. During these measurements we noticed that twin domains had reappeared in the $La_2CuO_4$ crystal, probably due to the experimental treatment like heating and cooling the sample which might result in strains induced *in situ* by magnetostriction. It was hence not possible to fully discriminate between situations in which the field is applied in the two in-plane directions *a* and *b*. The intensities of the magnetic reflections 100 and 102 were observed to decrease continuously in an in-plane field, while those at 101 and 103 increased (Fig. 4). This effect is fully reversible, and no hysteresis effects were observed. At 1.8 K, the 100 intensity decreases by about 20 % in an applied field of 14.5 T (Fig. 4). An extrapolation based on a power law behavior (see lines in fig. 4) yields a critical magnetic field of about 22 T. For higher temperatures, the critical field is reduced and reaches a value of 12.3 T at the temperature 300 K. This is close to the Néel temperature of 316(2) K (Fig. 2). These critical fields are comparable to, albeit somewhat higher than, those reported in magnetoresistance experiments on $La_{1.99}Sr_{0.01}CuO_4$. [10]

The field dependence of the magnetic intensities clearly suggests that the *C*-type ordering is continuously reduced with increasing field, and that a *G*-type ordering is induced. The moment direction of the induced order can be determined from the intensity ratio $I(101)/I(103)$. We found a ratio of about 3, which is consistent with a moment direction parallel to the *c*-axis. In contrast, models with moments parallel to *a* and *b* yield ratios of about 0.3 and 1, respectively, inconsistent with the data. The field-induced $G_z$ order we have found is in full agreement with the one inferred from Raman scattering work in magnetic fields along the *b*-axis. [11,12] The results of the representation analysis given in Table V show that the $G_z$ mode is compatible with the ferromagnetic mode $F_y$, which is too weak to be detectable directly in our neutron diffraction study. The lowering of the Zeeman energy along *b* allowed by this mode explains the origin of the field-induced spin reorientation and further supports the conclusions of Refs. 11 and 12. The full magnetic structure in a large *b*-axis oriented field is shown in Fig. 3.

The positions, intensities, and linewidths of nuclear Bragg peaks of the form *h0l* and *0kl* with *h,k,l* = even (allowed in the orthorhombic space group *Bmab*) did not show any field dependence up to 12 T. Unfortunately our cryomagnet did not allow us to reach magnetic fields exceeding 12 T at a sample temperature of 300 K. Therefore we were not able to probe whether the field-driven spin reorientation transition is related to the magnetic field induced detwinning of $La_2CuO_4$ observed by Lavrov *et al.* in a magnetic field of 14 T at room temperature. [9]

Fig. 5 shows the dependence of the intensities of the magnetic Bragg reflections 100 and 210 on a magnetic field applied parallel to the *c*-axis. At a field of $H_c$ = 11.5 T the 100 intensity decreases, whereas the 210 intensity increases. In contrast to the measurements in an in-plane field, an abrupt change of the intensity was observed, and this phase transition shows a hysteresis with a field width of about 1.5 T. This behaviour results from a field-induced $G_yF_z$ mode, as reported by Kastner *et al.* at a critical field of 4.8 T in a non-stoichiometric sample with a correspondingly reduced Néel temperature of 234 K. The high-field magnetic structure is depicted in Fig. 3. The origin of this transition was discussed at length in prior work. [14,24] Here we note briefly that the high-field spin structure is fully compatible with the representation analysis of Table V, and that the critical field is lower than that of the transition induced by the in-plane magnetic field, because the Zeeman energy competes against the exchange anisotropy as well as the interplanar exchange coupling.

## IV. CONCLUSIONS

In summary, our determination of the magnetic structure in high magnetic fields directly confirms the magnetic field-induced spin reorientations indicated by prior work on pure and lightly doped $La_2CuO_4$. These reorientations are driven by the interplay between the Zeeman energy, the weak exchange coupling between the $CuO_2$ layers, and the exchange anisotropies. The phenomena discussed here may also be of interest for models of magnetic correlations in superconducting $La_{2-x}Sr_xCuO_{4+y}$, because both the zero-field [26] and the magnetic-field-induced [27] incommensurate magnetic order recently observed in doped $La_2CuO_4$ exhibit at least short-range interplanar correlations as well as residual exchange anisotropies. We also discovered a subtle monoclinic distortion of the crystal structure of pure and lightly Sr-doped

La$_2$CuO$_4$, which results predominantly from displacements of the apical oxygen atoms. These displacements are small, but as the pinning of charge stripes in doped La$_2$CuO$_4$ appears to be determined by subtle crystallographic details [3], they may well be relevant for a complete understanding of the charge transport in this system as well.

## ACKNOWLEDGMENTS

We acknowledge the support of the Deutsche Forschungsgemeinschaft under grants UL 164/4 and KE923/1–2 in the consortium FOR538. We also acknowledge that the sample-preparation part of this work was supported by the Grant-in-Aid for Science provided by the Japan Society for the Promotion of Science.

Table I

Positional and thermal parameters of La$_2$CuO$_4$ at 325 K as obtained from the structure refinement in the orthorhombic and monoclinic space groups *Bmab* and *Bm*11, respectively. For the refinement in *Bm*11 the occupancies of the atoms were taken from the refinement in *Bmab* and were not allowed to vary.

La$_2$CuO$_4$: $a$ = 5.3568(6) Å, $b$ = 5.4058(5) Å, $c$ = 13.1432(11) Å.

| Atom | *Bmab* | x | y | z | $U_{11}$ | $U_{22}$ | $U_{33}$ | $U_{12}$ | $U_{13}$ | $U_{23}$ | occ |
|---|---|---|---|---|---|---|---|---|---|---|---|
| La | 8f | 0 | –0.00672(4) | 0.36145(2) | 0.71(2) | 0.62(1) | 0.44(1) | 0 | 0 | –0.04(1) | 1.0001(10) |
| Cu | 4a | 0 | 0 | 0 | 0.33(2) | 0.28(1) | 0.88(2) | 0 | 0 | 0.02(1) | 0.9991(15) |
| O1 | 8e | ¼ | ¼ | –0.00719(3) | 0.68(2) | 0.55(1) | 1.43(2) | –0.19(2) | 0 | 0 | 1.0016(14) |
| O2 | 8f | 0 | 0.03406(10) | 0.18367(3) | 2.08(3) | 1.57(2) | 0.66(2) | 0 | 0 | 0.09(1) | 0.9925(16) |

| Atom | *Bm*11 | x | y | z | $U_{11}$ | $U_{22}$ | $U_{33}$ | $U_{12}$ | $U_{13}$ | $U_{23}$ | occ |
|---|---|---|---|---|---|---|---|---|---|---|---|
| La1 | 2a | 0 | –0.00669(6) | 0.36149(2) | 0.73(2) | 0.62(1) | 0.45(1) | 0 | 0 | –0.02(1) | 1.0001 |
| La2 | 2a | 0 | 0.00669(6) | 0.63851(2) | 0.73(2) | 0.62(1) | 0.45(1) | 0 | 0 | –0.02(1) | 1.0001 |
| La3 | 2a | 0 | 0.50669(6) | 0.86149(2) | 0.73(2) | 0.62(1) | 0.45(1) | 0 | 0 | –0.02(1) | 1.0001 |
| La4 | 2a | 0 | 0.49331(6) | 0.13851(2) | 0.73(2) | 0.62(1) | 0.45(1) | 0 | 0 | –0.02(1) | 1.0001 |
| Cu1 | 4a | 0 | 0.0006(6) | –0.00156(6) | 0.28(2) | 0.29(2) | 0.86(2) | 0 | 0 | –0.02(1) | 0.9991 |
| Cu2 | 4a | ½ | 0.4996(5) | 0.00156(6) | 0.28(2) | 0.29(2) | 0.86(2) | 0 | 0 | –0.02(1) | 0.9991 |
| O11 | 4a | ¼ | 0.2497(7) | –0.00705(4) | 0.70(2) | 0.56(2) | 1.46(2) | –0.04(9) | –0.05(2) | –0.01(1) | 1.0016 |
| O12 | 4a | ¼ | 0.7496(7) | 0.00705(4) | 0.70(2) | 0.56(2) | 1.46(2) | –0.04(9) | –0.05(2) | –0.01(1) | 1.0016 |
| O21 | 2a | 0 | 0.0299(4) | 0.18365(3) | 2.06(3) | 1.52(2) | 0.69(2) | 0 | 0 | –0.00(2) | 0.9925 |
| O22 | 2a | 0 | –0.0380(4) | –0.18365(3) | 2.06(3) | 1.52(2) | 0.69(2) | 0 | 0 | –0.00(2) | 0.9925 |
| O23 | 2a | ½ | 0.5380(4) | –0.18365(3) | 2.06(3) | 1.52(2) | 0.69(2) | 0 | 0 | –0.00(2) | 0.9925 |
| O24 | 2a | ½ | –0.5299(4) | 0.18365(3) | 2.06(3) | 1.52(2) | 0.69(2) | 0 | 0 | –0.00(2) | 0.9925 |

The thermal parameters $U_{ij}$ (given in 100 Å$^2$) are in the form exp[$-2\pi^2(U_{11} h^2 a^{*2} + \ldots 2U_{13} h l a^* c^*)$].

Table II

Interatomic distances (Å) and angles (°) of La$_2$CuO$_4$ at 325 K.

|  | La$_2$CuO$_4$ |
|---|---|
| Cu−O1 | 1.9049(1) |
| Cu−O2 | 2.4210(5) |
| O1−Cu−O1 | 90.66(1) / 89.34(1) |
| La−O1 | 2.5885(4) |
| La−O2 | 2.3470(5) |
| La−O2 | 2.5523(7) |

Table III

Observed and calculated structure factors obtained from the structure refinements in the monoclinic space group *Pm* (*Pm*11). In the Table only the structure factors are listed that are systematically zero in *Bmab*.

| h k l | $F_{obs}^2$ | $F_{cal}^2$ | h k l | $F_{obs}^2$ | $F_{cal}^2$ | h k l | $F_{obs}^2$ | $F_{cal}^2$ |
|---|---|---|---|---|---|---|---|---|
| 2 1 0 | 0.25 | 0.11 | 3 0 1 | 0.29 | 0.03 | 1 0 9 | 0.21 | 0.24 |
| 4 1 0 | 0.21 | 0.09 | 5 0 1 | 0.28 | 0.02 | 3 0 9 | 0.35 | 0.27 |
| 6 1 0 | 0.09 | 0.07 | 7 0 1 | 0.20 | 0.03 | 5 0 9 | 0.23 | 0.20 |
| 0 3 0 | 0.30 | 0.25 | 1 0 3 | 0.28 | 0.09 | 1 0 11 | 0.22 | 0.28 |
| 2 3 0 | 0.21 | 0.24 | 3 0 3 | 0.31 | 0.10 | 3 0 11 | 0.22 | 0.32 |
| 4 3 0 | 0.25 | 0.20 | 5 0 3 | 0.20 | 0.07 | 5 0 11 | 0.22 | 0.23 |
| 6 3 0 | 0.17 | 0.15 | 7 0 3 | 0.21 | 0.10 | 1 0 13 | 0.25 | 0.32 |
| 0 5 0 | 0.24 | 0.20 | 1 0 5 | 0.32 | 0.14 | 3 0 13 | 0.26 | 0.36 |
| 2 5 0 | 0.18 | 0.19 | 3 0 5 | 0.28 | 0.16 | 5 0 13 | 0.28 | 0.27 |
| 4 5 0 | 0.20 | 0.16 | 5 0 5 | 0.18 | 0.11 | 1 0 15 | 0.31 | 0.35 |
| 6 5 0 | 0.16 | 0.16 | 7 0 5 | 0.25 | 0.17 | 3 0 15 | 0.34 | 0.39 |
| 0 7 0 | 0.02 | 0.02 | 1 0 7 | 0.37 | 0.19 | 5 0 15 | 0.26 | 0.30 |
| 2 7 0 | 0.02 | 0.02 | 3 0 7 | 0.30 | 0.22 | 1 0 17 | 0.29 | 0.38 |
| 4 7 0 | 0.02 | 0.01 | 5 0 7 | 0.25 | 0.16 | 1 0 19 | 0.29 | 0.39 |
| 1 0 1 | 0.34 | 0.03 | 7 0 7 | 0.24 | 0.23 | 1 0 21 | 0.28 | 0.40 |

Table V. Representations of the base vectors of {Cu} in La$_2$CuO$_4$ with the space group *Bmab*. The copper atoms are located at the Wyckoff position 4*a*: 0,0,0 (1), 0,½,½ (2), ½,0,½ (3) and ½,½,0 (4). The modes are $F(+ + + +)$, $C(+ + - -)$, $G(+ - - +)$ and $A(+ - + -)$.

| Bmab | x | y | z |
|---|---|---|---|
| $\Gamma_1$ | $G_x$ | – | – |
| $\Gamma_2$ | – | $G_y$ | $F_z$ |
| $\Gamma_3$ | $F_x$ | – | – |
| $\Gamma_4$ | – | $F_y$ | $G_z$ |
| $\Gamma_5$ | $C_x$ | – | – |
| $\Gamma_6$ | – | $C_y$ | $A_z$ |
| $\Gamma_7$ | $A_x$ | – | – |
| $\Gamma_8$ | – | $A_y$ | $C_z$ |

Table VI. Representations of the base vectors of {Cu} in La$_2$CuO$_4$ with the space group *B*2/*m*. The copper atoms are located at the Wyckoff position 2*a* [in 0,0,0 and ½,0,½] and 2*d* [in ½,½,0 and 0,½,½. The modes are $F(+ +)$ and $A(+ -)$.

| B2/m | x | y | z |
|---|---|---|---|
| $\Gamma_1$ | $F_x$ | – | – |
| $\Gamma_2$ | $A_x$ | – | – |
| $\Gamma_3$ | – | $A_z$ | $A_z$ |
| $\Gamma_4$ | – | $F_y$ | $F_z$ |

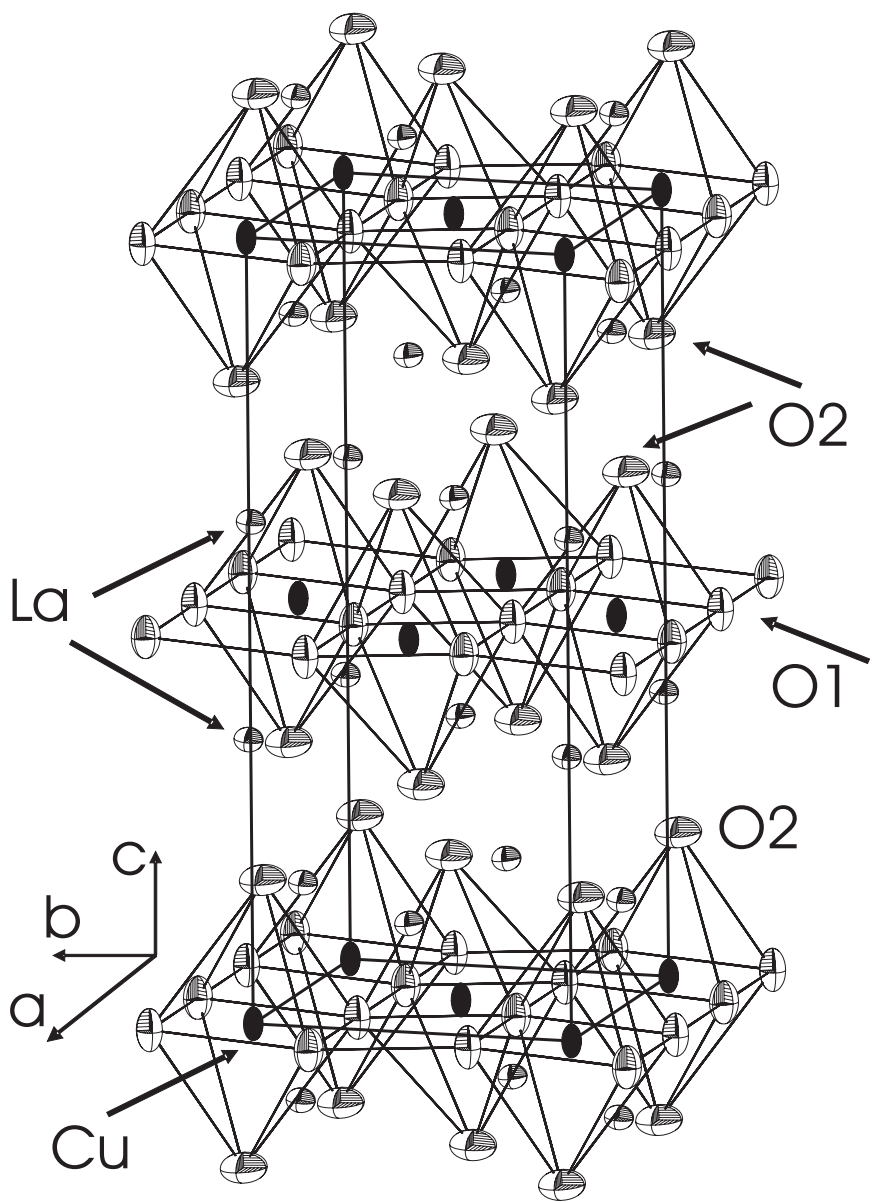

Fig. 1. Crystal structure of $La_2CuO_4$ at 325 K. Shown are the distorted corner-shared $CuO_6$-octahedra in the *ab*-plane. The thermal ellipsoids of the atoms are plotted at the 90 % probability level.

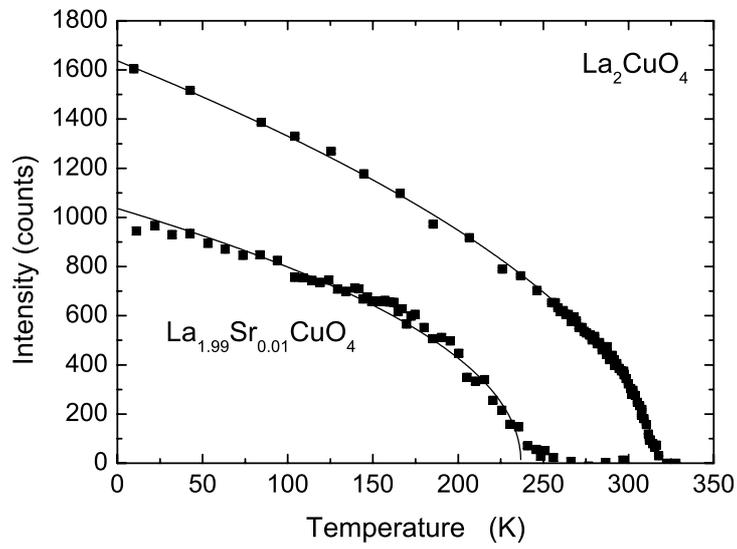

Fig. 2. Temperature dependence of the intensity of the 100 magnetic Bragg reflections of $La_2CuO_4$ and $La_{1.99}Sr_{0.01}CuO_4$.

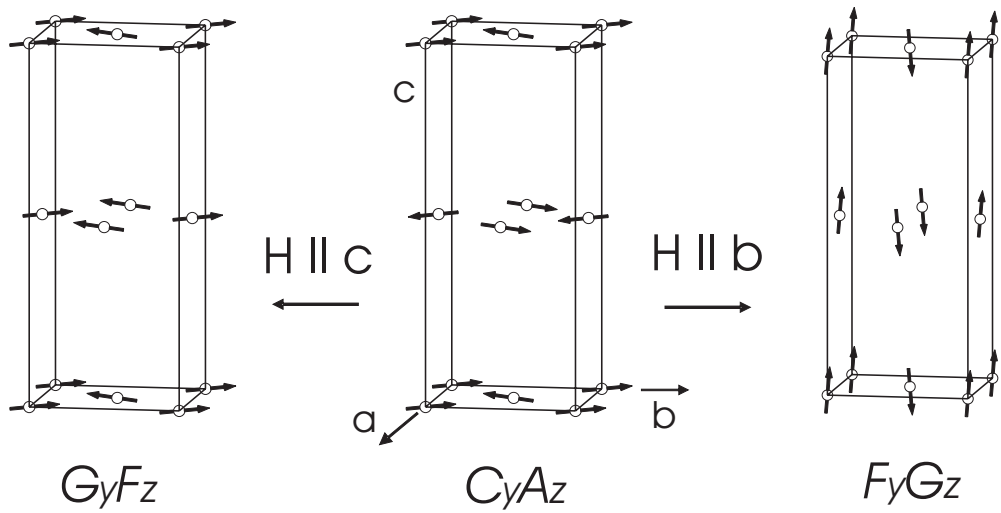

Fig. 3. Magnetic structure types of $La_2CuO_4$. Shown are the magnetic structure in zero-field (middle) and applied magnetic field parallel to the *b*- and *c*-axis. The components with the modes $F_z$ ($H // c$), $A_z$ ($H = 0$) and $F_y$ ($H // b$) are rather weak. For clarity, they are exaggerated in the figures.

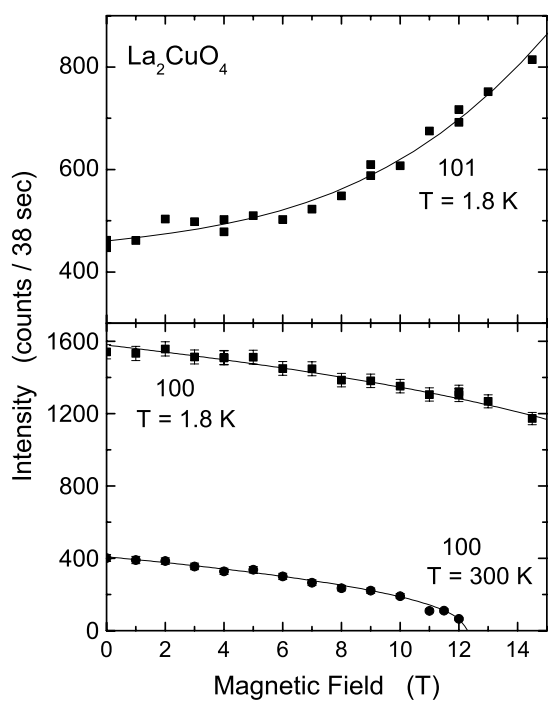

Fig. 4. Magnetic field dependence of the magnetic Bragg reflections 100 and 101 of $La_2CuO_4$ for a magnetic field applied within the *ab*-plane.

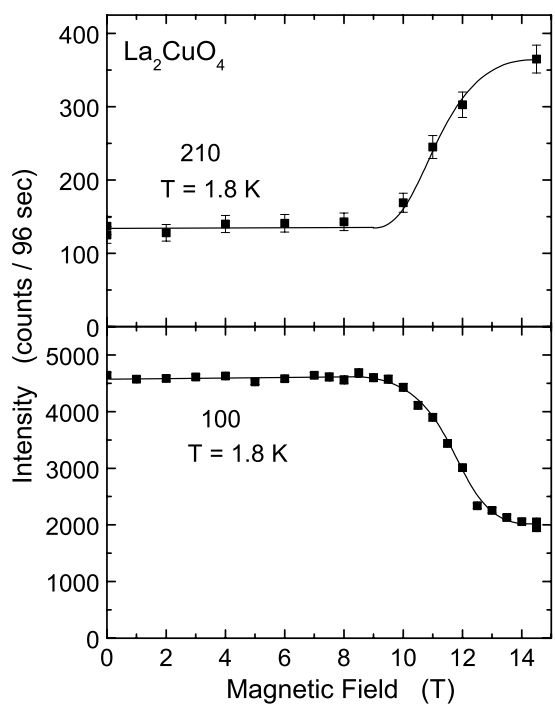

Fig. 5. Magnetic field dependence of the magnetic reflections 100 and 210 of $La_2CuO_4$ for a magnetic field applied parallel to the crystallographic *c*-axis, measured at $T = 1.8$ K. Lines are a guide to the eye.